# Tunable superconductivity coexisting with the anomalous Hall effect in 1$T'$-WS$_2$

**Authors:** Md Shafayat Hossain[1]*†, Qi Zhang[1]*, David Graf[2]*, Mikel Iraola[3,4]*, Tobias Müller[5], Sougata Mardanya[6], Yi-Hsin Tu[7], Zhuangchai Lai[8,9], Martina O. Soldini[5], Siyuan Li[8], Yao Yao[8], Yu-Xiao Jiang[1], Zi-Jia Cheng[1], Maksim Litskevich[1], Brian Casas[2], Tyler A. Cochran[1], Xian P. Yang[1], Byunghoon Kim[1], Kenji Watanabe[10], Takashi Taniguchi[11], Sugata Chowdhury[6], Arun Bansil[12], Hua Zhang[8,13,14], Tay-Rong Chang[5], Mark Fischer[5], Titus Neupert[5], Luis Balicas[2], M. Zahid Hasan[1,15]†

**Affiliations:**

[1]Laboratory for Topological Quantum Matter and Advanced Spectroscopy, Department of Physics, Princeton University, Princeton, New Jersey, USA.

[2]National High Magnetic Field Laboratory, Tallahassee, Florida 32310, USA.

[3]Donostia International Physics Center, 20018 Donostia-San Sebastian, Spain.

[4]Institute for Theoretical Solid State Physics, IFW Dresden, Helmholtzstrasse 20, Dresden, Germany.

[5]Department of Physics, University of Zurich, Winterthurerstrasse, 190, 8057 Zurich, Switzerland.

[6]Department of Physics and Astrophysics, Howard University, Washington, USA.

[7]Department of Physics, National Cheng Kung University, 701 Tainan, Taiwan.

[8]Department of Chemistry, City University of Hong Kong, Hong Kong, China.

[9]Department of Applied Physics, The Hong Kong Polytechnic University, Hong Kong, China.

[10]Research Center for Electronic and Optical Materials, National Institute for Materials Science, 1-1 Namiki, Tsukuba 305-0044, Japan.

[11]Research Center for Materials Nanoarchitectonics, National Institute for Materials Science, 1-1 Namiki, Tsukuba 305-0044, Japan.

[12]Department of Physics, Northeastern University, Boston, Massachusetts 02115, USA.

[13]Hong Kong Branch of National Precious Metals Material Engineering Research Center (NPMM), City University of Hong Kong, Hong Kong, China.

[14]Shenzhen Research Institute, City University of Hong Kong, Shenzhen 518057, China.

[15]Quantum Science Center at ORNL, Oak Ridge, TN, USA.

* These authors contributed equally to this work

†Corresponding to: mdsh@princeton.edu; mzhasan@princeton.edu

## Abstract:

**Transition metal dichalcogenides are a family of quasi-two-dimensional materials that display a high technological potential due to their wide range of electronic ground states, e.g., from superconducting to semiconducting, depending on the chemical composition, crystal structure, or electrostatic doping. Here, we unveil that by tuning a single parameter, the hydrostatic pressure $P$, a cascade of electronic phase transitions can be induced in the few-layer transition metal dichalcogenide $1T'$-$WS_2$, including superconducting, topological, and anomalous Hall effect phases. Specifically, as $P$ increases, we observe a dual phase transition: the suppression of superconductivity with the concomitant emergence of an anomalous Hall effect at $P \approx$ 1.15 GPa. Remarkably, upon further increasing the pressure above 1.6 GPa, we uncover a reentrant superconducting state that emerges out of a state still exhibiting an anomalous Hall effect. This superconducting state shows a marked increase in superconducting anisotropy with respect to the phase observed at ambient pressure, suggesting a different superconducting state with a distinct pairing symmetry. Via first-principles calculations, we demonstrate that the system concomitantly transitions into a strong topological phase with markedly different band orbital characters and Fermi surfaces contributing to the superconductivity. These findings position $1T'$-$WS_2$ as a unique, tunable superconductor, wherein superconductivity, anomalous transport, and band features can be tuned through the application of moderate pressures.**

## Main Text:

Tunable electronic[1], magnetic[2], and ferroelectric[3] properties in two-dimensional materials, e.g., as a function of electric field, electrostatic doping, pressure, number of layers, or relative twist angle, not only facilitate studying the fundamentals of phase transitions in the two-dimensional limit but also unveil their potential for applications. In this context, two-dimensional moiré materials[4-20], including twisted bilayer graphene[4-10] and twisted transition metal dichalcogenides[11-16], have garnered significant recent attention due to the tunability of their physical response as a function of the twist angle and electrostatic gating, which leads to a myriad of ground states including topological phases like the Chern insulators[17], integer[10] and fractional quantum anomalous Hall effects[20] and strongly correlated ones such as Mott insulators[5,11], Wigner crystals[11], magnetism[9], and unconventional superconductivity[6]. Here, we unveil the remarkable tunability of the electronic and superconducting properties of a seemingly simple topological metal: atomically thin $1T'$-$WS_2$. We reversibly tune this system into a regime that simultaneously displays an anomalous Hall effect, usually the hallmark of intrinsic broken time-reversal symmetry and superconductivity, suggesting an unconventional superconducting pairing symmetry.

At ambient pressure $1T'$-$WS_2$ already stands as a promising candidate for topological superconductivity[21-25] since, as we discuss below, its band structure is characterized by topologically non-trivial indices. Its close relative, $2M$-$WS_2$, which has undergone more extensive studies, harbors topological surface states[26,27] as well as Majorana bound states within its vortex cores[28,29]. Additionally, $2M$-$WS_2$ was identified as an exotic spin-orbit-parity coupled superconductor[25], responsible for features such as a two-fold symmetric superconducting state and an exceptionally high upper critical magnetic field, *i.e.*, that surpasses the Pauli paramagnetic limit. These unique attributes have spurred immense interest in transition metal dichalcogenides at the intersection of topology and superconductivity[21,22]. However, $1T'$-$WS_2$ poses a unique challenge due to its metallic nature characterized by a high carrier density[21,22]. The large carrier density makes the precise control of its electronic state via electrostatic gating quite challenging. In contrast, we find that the mechanical manipulation of its lattice parameters provides a fruitful avenue for reversibly tuning the electronic properties without introducing impurities in $1T'$-$WS_2$. In this study, we harness hydrostatic pressure as a means for tuning the lattice parameters of $1T'$-$WS_2$ and hence, its electronic band structure. Through systematic pressure-dependent electrical transport experiments, we observe the emergence of the anomalous Hall effect upon increasing the pressure, which coincides with the suppression of the superconducting state observed at ambient pressure. Intriguingly, as we continue to increase pressure, while remaining within the metallic state characterized by the anomalous Hall effect, we observe the emergence of a pressure-induced reentrant superconducting phase displaying very high upper critical fields, *i.e.*, 2.5 times larger than the weak coupling Pauli limiting field value.

To conduct transport measurements under various hydrostatic pressures, we fabricated Hall bar devices using mechanically exfoliated few-layer (twelve-layers) $1T'$-$WS_2$ (see Fig. 1**a**-1**d** and Methods Section I) and positioned the sample within a piston-cylinder pressure cell, using Daphne oil as the pressure medium. The highlight of our experiment is illustrated by Figs. 1**e**-1**g** displaying the superconducting phase diagram as a function of the hydrostatic pressure based on data collected from three distinct samples. Figure 1**e** portrays the critical temperature ($T_c$) of the superconducting transition plotted against the external hydrostatic pressure ($P$). Initially, $T_c$ exhibits a gradual decline with increasing $P$, followed by its abrupt vanishing at $P = 1.15$ GPa, implying the suppression of superconductivity (referred to as SC1). Upon further increasing $P$, a second superconducting dome (noted as SC2) emerges at $P = 1.8$ GPa, which is characterized by a lower $T_c$ with respect to SC1. Next, we investigate the upper critical magnetic fields, which we define as the field at which the sample resistance drops to half of its value in the normal state just above the transition. When comparing SC2 to SC1, we find that the out-of-plane superconducting upper critical magnetic fields ($\mu_0 H_{c2\perp bc}$), for fields perpendicular to the *bc*-plane, and the ratio $\mu_0 H_{c2\perp bc}/T_c$ are markedly smaller (Fig. 1**f**) than the values measured for SC1. Surprisingly, the in-plane upper critical magnetic fields ($\mu_0 H_{c2||bc}$) for the SC2 phase for fields oriented along the *c*-axis exhibit similar magnitudes relative to those of the SC1 state, leading to a twofold increase in the ratio $\mu_0 H_{c2||c}/T_c$ for the SC2 state (Fig. 1**g**). This distinct behavior for the in-plane and out-of-plane upper critical fields reveal a substantial increase in the superconducting anisotropy $\gamma$, defined as the ratio $\gamma = H_{c2\|bc}/H_{c2\perp bc} = \xi_{\|bc}/\xi_{\perp bc}$, where $\xi_{\|bc}$ and $\xi_{\perp bc}$ represent the coherence lengths along and perpendicular to the *bc*-plane, respectively. This remarkable increase in anisotropy indicates that the superconducting properties of the SC2 state differ from those of SC1 (see also Extended Fig. 1), *i.e.*, characterized by a larger in-plane coherence length versus a decreased inter-planar one.

Figure 2 provides illustrative examples of the raw electrical transport data, from which we constructed the phase diagrams depicted in Fig. 1**e**-**g**. The temperature ($T$) dependence of the four-probe resistance ($R$) of a $1T'$-$WS_2$ sample reveals a metallic response under all pressures (shown in Fig. 2**a**). At ambient pressure ($P \cong 0$), the $R$ as a function of $T$ plot demonstrates a discernible superconducting transition. With increasing pressure, this transition steadily shifts to lower temperatures, a trend that becomes evident by comparing the traces corresponding to pressures of 0.1, 0.5, and 1.0 GPa (as shown in Fig. 2**a**), until it vanishes at $P = 1.15$ GPa. The analysis of the magnetic-field dependence, conducted for fields aligned both perpendicular to the *bc*-plane and along the *c*-axis, reveals a consistent pattern: the superconducting transition shifts to lower fields within the SC1 state as pressure increases (Fig. 2**b**,**c**). In an intermediate pressure range, exemplified by the traces collected under $P = 1.15$, 1.5, and 1.63 GPa in Fig. 2**a**,**b**, no superconducting transition is observed. Remarkably, at $P = 1.8$ GPa, the superconducting transition resurfaces with a notably lower transition temperature (Fig. 2**a**) and reduced upper critical magnetic field for fields perpendicular to the *bc*-plane (Fig. 2**b**). However, for fields parallel to the *bc*-plane, the transition displays similar upper critical fields relative those of the SC1 state. For instance, the trace at $P = 2.3$ GPa (Fig. 2**c**) illustrates this occurrence, resulting in substantial values for the ratios $\mu_0 H_{c2||bc}/T_c$. Notice that our transport results are reproducible upon depressurization, indicating a return to the original crystallographic structure. Extended Fig. 2 presents transport data at 0.1 GPa, showing measurements taken as pressure increased from 0 to 0.1 GPa and then decreased from 2.3 GPa to 0.1 GPa. Both measurements display the same transport behavior, demonstrating reversibility.

After uncovering the superconducting phase diagram of 1*T*′-WS$_2$, we turn to investigating its normal state. To achieve this, we conducted Hall effect measurements (depicted for $T = 10$ K in Fig. 3 and Extended Fig. 2) for various pressures. These measurements provide crucial insights into the distinctions between the SC1 and SC2 states, by shedding light on the normal state from which superconductivity nucleates. When examining the Hall resistance within the pressure range where the SC1 state is observed, a linear response as a function of the field is observed (Fig. 3**a** and 3**b** for $P = 0$ and 1.0 GPa, respectively). This is consistent with prior findings under vacuum[21,22]. Upon further increasing pressure, a noteworthy development occurs—precisely at $P = 1.15$ GPa, where the superconducting state vanishes, an anomalous Hall component emerges (Extended Fig. 3**a**,**b**). The anomalous Hall effect is also present at pressures $P > 1.15$ GPa, as evident in the $P = 1.63$ GPa data (Fig. 3**c**,**d**). Surprisingly, even at higher pressures where SC2 is observed the normal state continues to exhibit an anomalous Hall effect, albeit less pronounced; refer to Fig. 3**e**,**f** ($P = 2.3$ GPa) and Extended Fig. 2**c**,**d** ($P = 1.8$ GPa), where the anomalous Hall effect is evident but less prominent relative to the signal observed in the range $1.15 < P < 1.8$ GPa. It is worth noting that no hysteresis is observed in these experiments.

Lastly, we examine the temperature dependence of the anomalous Hall effect. Figure 4 showcases the key data, where we provide two examples: one at $P = 1.63$ GPa, a pressure where no superconductivity is observed at low temperatures, and the other at $P = 2.3$ GPa, where the reentrant superconductivity becomes prominent. In both instances, a strong temperature dependence is observed for the anomalous component, while the linear-in-field component exhibits minimal sensitivity to temperature. Notably, at $P = 1.63$ GPa, where the anomalous Hall effect is more pronounced, this effect significantly diminishes at a temperature of $T = 28$ K (Fig. 4**a**,**b**). Conversely, at $P = 2.3$ GPa, the anomalous Hall effect vanishes entirely by $T = 25$ K (Fig. 4**c**,**d**). This temperature-dependent behavior provides an indicative energy scale for the destabilization of the underlying magnetic order. Notably, the *R* as a function of *T* data (Extended Fig. 4) reveals a weak anomaly near $T = 30$ K, which nearly coincides with the disappearance of the anomalous Hall effect. A clear understanding of what this anomaly represents and whether it signals a phase-transition would require structural analysis under high pressures and therefore will be left for future research.

Examining the temperature-dependent anomalous Hall effect ($R_{xy}^{AHE}$) data at these two pressures hints at a potential competition between the anomalous Hall effect phase and superconductivity. First, at a given temperature, the saturated value of $R_{xy}^{AHE}$ at $P = 2.3$ GPa (where the reentrant superconducting state is present) is lower than the one measured at $P = 1.63$ GPa, where the sample does not exhibit superconductivity. Second, at $P = 2.3$ GPa, and below $T_c$, the saturated value of $R_{xy}^{AHE}$ appears to decrease slightly compared to its immediate higher-temperature value. This contrasts with the behavior measured at $P = 1.63$ GPa, where the saturated value of $R_{xy}^{AHE}$ consistently increases as the temperature is lowered. These observations indicate a possible competition between superconductivity and the anomalous Hall effect. The reentrant superconducting state manifests when $R_{xy}^{AHE}$ is suppressed at low temperatures.

The wealth of pressure-tunable electronic states in 1$T'$-WS$_2$, as depicted in Figs. 1-4, prompts us to elucidate the dependence of the low-energy electronic structure on the applied pressure (see Methods and Extended Figs. 5, 6 and Extended data Table 1, 2 for details). To this end, we performed a first-principles calculations of the electronic band structure as a function of pressure, which allows us to track changes occurring around the Fermi energy as a function of pressure, as well as identify possible topological phase transitions (Fig. 5**a**,**d**-**f**). For our calculations, we assume that 1$T'$-WS$_2$ remains structurally stable under pressure. This assumption is supported by two key observations: (*i*) our transport measurements reveal a reversible pressure evolution (Extended Fig. 2), and (*ii*) our phonon calculations show no signs of phononic instability within the studied pressure range (Extended Fig. 7). We observe that the system undergoes a topological phase transition upon increasing pressure, where the topological indices are defined with respect to the direct gap above the band leading to electron-hole compensation: at $P = 0$ GPa, the system realizes a topological crystalline insulator followed, at $P \sim 1.0$ GPa, by a phase transition as a result of a band inversion at the point $Y$, which drives the system into a strong topological phase. We performed a slab calculation at $P = 2.3$ GPa (Fig. 5**c**), therefore in the strong topological phase, finding that the surface perpendicular to the stacking direction exhibits a cone at $\bar{\Gamma}$ located 50 meV above the Fermi level. Since these boundary modes reside at higher energies, they are not expected to influence the low-energy transport properties of the system but may be accessed by other experimental probes.

The change in topology, however, has a profound effect on the orbital makeup of the bands at the Fermi surface. In the low-pressure phase, these bands exhibit a mixed *d*- and *p*-orbital nature around *Y*, whereas in the strong topological phase, their character changes to predominantly *p*-orbital. It would be interesting to further explore how this orbital transformation is linked to the breakdown of superconductivity. At lower energies, we find the emergence of a markedly three-dimensional topography for the Fermi surface, despite the layered nature of the material. At $P = 0$ GPa (Fig. 5**d**), the Fermi level is fixed at perfect electron-hole compensation, which is in line with our Hall transport measurements, revealing linear behavior as a function of the magnetic field at low temperatures, despite the presence of both types of carriers. With increasing pressure, we first observe a Lifshitz transition merging both electron-like Fermi surfaces (Fig. 5**e**). Subsequently, at yet higher values of pressure, we observe merged Fermi surface sheets to cross the Brillouin zone boundaries in addition to the appearance of a hole in the Fermi sheet around the *Y* point (Fig. 5**f**). Lifshitz transitions are connected to van Hove singularities in two dimensions, which are singularities in the density of states, and can introduce new possible pairing momenta, thus marking potential 'hot spots' on the Fermi surface for nucleating superconductivity.

Having elucidated the electronic structure of 1$T'$-WS$_2$ under pressure, we discuss the plausible explanations for the experimental observations. A possible origin for the emergence of the anomalous Hall effect, particularly in strongly spin-orbit coupled systems such as 1$T'$-WS$_2$, is magnetic order. However, a sizable anomalous Hall effect has also been observed in paramagnetic systems, such as ZnO/MnZnO[30], or Kagome metals where some form of fluctuating, or weak magnetism is debated, but conclusive evidence for long-range magnetic order is lacking[31]. Another possible origin for such an anomalous Hall effect is skew scattering on localized magnetic moments. However, S is not magnetic while W is known to be paramagnetic. Furthermore, anomalous Hall is not observed at lower pressures, indicating the absence of magnetic impurities. Still, the significant temperature dependence of the anomalous Hall signal, with a near-complete attenuation of the anomalous Hall effect around $T = 30$ K, hints towards an itinerant nature of the magnetic order relating to electron-electron interactions. To elucidate this possibility theoretically, we derived Wannierized tight-binding models at fixed values of pressure $P = 0$ GPa and $P = 2.3$ GPa and used the random phase approximation (RPA) method to evaluate the susceptibility in the system (Fig. 5**b**). The RPA results allow us to identify possible charge and spin instabilities of the Fermi liquid. Specifically, we obtain a leading divergence of the susceptibility at a star of incommensurate wave vectors (Fig. 5**b**), which can either lead to a single-$q$ or multi-$q$ ordered state. In addition, this leading divergence is magnetic in nature. Therefore, it may lead to an incommensurate spin density wave. Such a magnetic tendency, and associated magnetic fluctuations, offer a plausible explanation for the observed anomalous Hall response[32]. Therefore, our discovery of the anomalous Hall effect in $1T'$-WS$_2$ encourages further investigations into magnetic order using complementary experimental techniques, such as muon spin relaxation experiments.

Upon delineating the plausible origins of the observed anomalous Hall effect, it is worth discussing how it may influence the reentrant superconducting state. In fact, our experimental findings clearly indicate a correlation between the superconductivity and the anomalous Hall effect. First, the SC1 state disappears when the anomalous Hall effect emerges. The anomalous Hall effect is prominent in the pressure range where superconductivity is absent and becomes relatively subdued in the pressure range where the reentrant superconducting state (SC2) emerges. Conversely, $T_c$ is higher in the SC1 state where the anomalous Hall effect is not observed, while it is lower in the SC2 state where the anomalous Hall effect is observed. These observations also raise the possibility of an unconventional, perhaps a spin-triplet paired, superconducting state in phase SC2, which prompts us to discuss plausible order parameters for the superconducting phases within $1T'$-WS$_2$, leveraging a symmetry analysis. The low symmetry inherent to the $1T'$ structure limits the potential irreducible representations for the superconducting order parameter to four: $A_g$, $B_g$, $A_u$, and $B_u$, as dictated by the $C_{2h}$ point group symmetry. This group is generated by inversion symmetry and a (nonsymmorphic) $C_2$ symmetry or — equivalently — to an in-plane mirror symmetry. Among these representations, $A_g$ and $B_g$ ($A_u$ and $B_u$) are characterized as even (odd) under inversion, whereas $A_g$ and $B_u$ ($A_u$ and $B_g$) belong to mirror-even (mirror-odd) categories. Notably, all these irreducible representations are one-dimensional. This framework leads us to discount the likelihood of spontaneous time-reversal symmetry breaking stemming from the superconducting order in $1T'$-WS$_2$ – a mechanism seen in chiral $p$-wave superconductors. Nonetheless, it is plausible for superconductivity to emerge in a phase, where time-reversal symmetry is broken by an alternative mechanism. Neglecting multiorbital effects, we can associate $A_g$ and $B_g$ with singlet pairing and $A_u$ and $B_u$ with triplet pairing. Should we assume that the ambient pressure scenario realizes the $A_g$ phase (fully gapped), and that the reentrant superconducting phase unveiled in this study is distinct, three potential candidates remain for its order parameter: $A_u$, $B_g$, and $B_u$. Among these, both $B_g$ and $A_u$ exhibit nodal characteristics, featuring line nodes in three dimensions and point nodes in two dimensions. Their symmetry character manifests as $d_{xy}$ and $p_x$, respectively. In contrast, $B_u$ may potentially be nodeless. In a time-reversal symmetry broken environment, suggested by the anomalous Hall effect observed with SC2, we can speculate that the triplet phases $A_u$ and $B_u$ might be more inherently compatible with this reentrant superconducting state. The possibility of an unconventional superconducting phase at re-entrance would be consistent with our first-principles calculations, which point to a progressive weakening of the electron-phonon coupling with the concomitant increase in the spin susceptibility as the hydrostatic pressure increases (Extended Figs. 7-9).

Superconducting phases in WS$_2$ were previously studied under hydrostatic pressure. For example, in the $2M$-WS$_2$ phase, the superconducting state gradually weakens upon increasing pressure[33], while in the $3R$-WS$_2$ phase, superconductivity only emerges at pressures exceeding 45 GPa[34]. Insofar, there are no reports on the effect of pressure on the superconductivity of the $1T'$ phase of WS$_2$. The $1T'$ phase we examine here under hydrostatic pressure features subtle differences in unit cell parameters (see Supplementary Information) with respect to the other previously studied WS$_2$ phases. These structural differences are likely to contribute to significant differences in the observed high-pressure behavior with respect to the other phases, as exposed by our experiments.

In conclusion, we have discovered pressure-induced reentrant superconductivity coexisting with the anomalous Hall effect in atomically thin 1$T'$-WS$_2$. The system undergoes a series of phase transitions under increasing pressure, as depicted schematically in Extended Fig. 10. The superconducting state gradually weakens with increasing pressure, indicated by diminishing $T_c$ and $H_{c2}$ values. This transition progresses with the suppression of $T_c$ at 1.15 GPa coinciding with the onset of the anomalous Hall effect. Interestingly, at $P = 1.8$ GPa, a striking "Lazarus effect" unfolds, wherein superconductivity reemerges, albeit with a lower $T_c$ and an increased superconducting anisotropy while still coexisting with a less pronounced anomalous Hall effect. This suggests an interplay between two distinct electronic orders, *i.e.*, superconductivity with magnetism. The reentrant superconducting state exhibits substantially elevated $\mu_0 H_{c2\|bc}/T_c$ values, while $\mu_0 H_{c2\perp bc}/T_c$ decreases, indicating a superconducting state that is distinct from the one at low pressures, possibly of an unconventional nature. This discovery represents clear evidence for the pressure tuning of the electronic properties of atomically thin 1$T'$-WS$_2$, manifesting as pressure-induced anomalous transport and potentially, an unconventional superconducting phase. Furthermore, with our findings, 1$T'$-WS$_2$ joins a rare but growing number of materials (for instance, the AV$_3$Sb$_5$ Kagomes; A≡ K, Rb, Cs)[35], in which superconductivity nucleates in a phase exhibiting signatures of broken time-reversal-symmetry in the form of an anomalous Hall-effect. However, the consequences of time-reversal-symmetry breaking for the superconducting pairing symmetry in 1$T'$-WS$_2$ remains unclear and as an open question for future research.

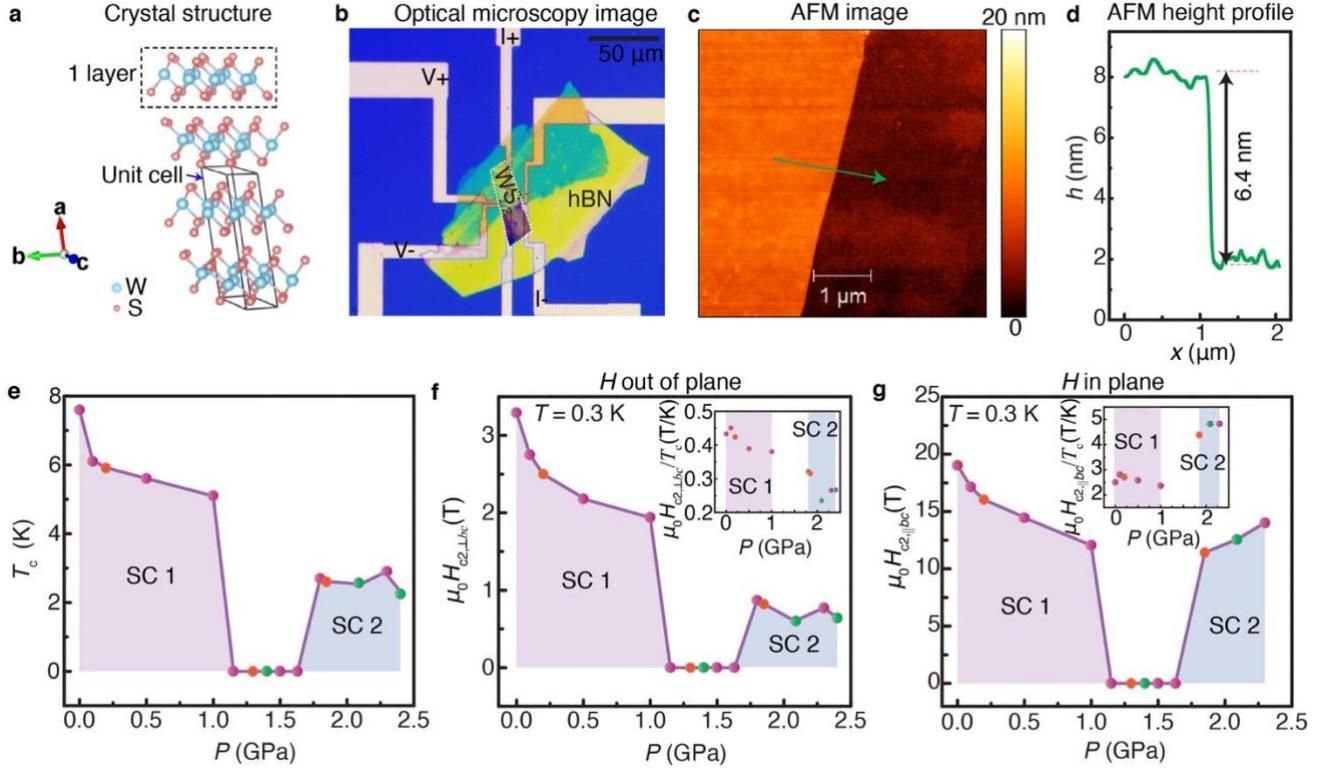

**Fig. 1: Observation of a tunable superconducting state in atomically thin 1T′-WS$_2$. a,** Crystal structure and the unit cell of 1T′-WS$_2$ crystal. **b,** Optical microscopy image of an atomically thin 1T′-WS$_2$ flake, obtained through mechanically exfoliation from the crystal's *bc* plane and subsequently encapsulated with hBN. The flake, comprising approximately twelve atomic layers, is demarcated by white dashed lines. Current (I$^+$ and I$^-$) and voltage (V$^+$ and V$^-$) probes for four-terminal electrical transport measurements are also marked. **c,** Atomic force microscopy image of a 1T′-WS$_2$ flake with the same optical contrast as the flake in panel **b**. **d,** Atomic force microscopy height (*h*) profile collected along the green line in panel **c** (the arrow indicates the directions of the scan), revealing a thickness of $h \simeq 6.4$ nm that aligns with the expected thickness for twelve atomic layers of 1T′-WS$_2$. **e,** Critical temperature ($T_c$) of the superconducting transition plotted against the external hydrostatic pressure (*P*). Initially, $T_c$ decreases with increasing pressure, leading to the disappearance of the superconducting state SC1, at $P = 1.15$ GPa. However, at $P = 1.8$ GPa, a new superconducting state (SC2) emerges with a smaller $T_c$ relative to the first superconducting state, SC1. **f,** Out-of-plane upper critical magnetic field ($\mu_0 H_{c2\perp bc}$) for fields oriented perpendicular to the *bc* plane, defined as the field at which the resistance becomes half of its value in the normal state, plotted against pressure. Akin to $T_c$, $\mu_0 H_{c2\perp bc}$ decreases with increasing *P* in the SC1 state and then

resuscitates with a smaller value in the SC2 state. The inset shows the $P$ dependence of $\mu_0 H_{c2\perp bc}$ normalized by $T_c$, with $\mu_0 H_{c2\perp bc}/T_c$ being smaller in the SC2 state. **g**, Pressure-dependence of the in-plane upper critical magnetic field ($\mu_0 H_{c2||bc}$) for fields directed along the *bc* plane. Like $T_c$ and $\mu_0 H_{c2\perp bc}$, $\mu_0 H_{c2||bc}$ decreases with increasing $P$ in the SC1 state. However, after the reemergence of superconductivity, $\mu_0 H_{c2||bc}$ in the SC2 state becomes similar in magnitude to that of the SC1 state, despite the comparatively much smaller $T_c$ and $\mu_0 H_{c2\perp bc}$. The inset illustrates the $P$ dependence of $\mu_0 H_{c2||bc}$ normalized by $T_c$, revealing a nearly two-fold larger $\mu_0 H_{c2||bc}/T_c$ in the SC2 state. Data in panels **e-g** were obtained from three samples with identical thicknesses which exhibit similar characteristics, and are represented by the purple, green, and orange data points.

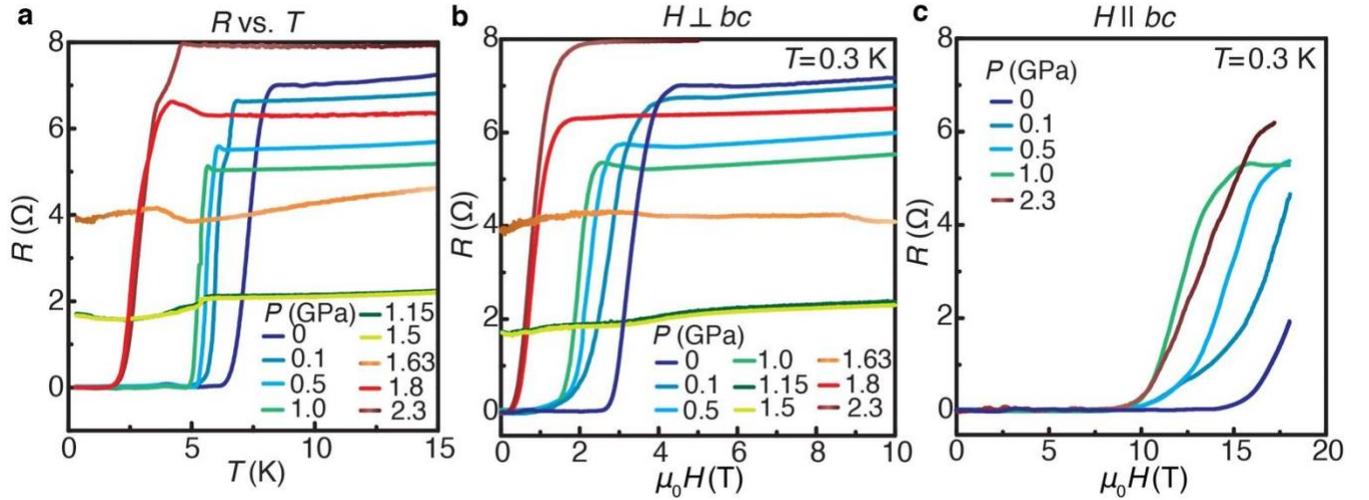

**Fig. 2: Transport measurements under several pressures revealing two superconducting states. a**, Temperature ($T$) dependence of the four-probe resistance ($R$) for the $1T'$-$WS_2$ device measured at different hydrostatic pressures ($P$). The sample exhibits metallic behavior at all pressures. At $P = 0$, $R$ as a function of $T$ shows a superconducting transition, which shifts to lower temperatures with increasing pressure before vanishing at $P = 1.15$ GPa. However, superconductivity reappears at 1.8 GPa in the reentrant state SC2, occurring at notably lower temperatures relative to the first superconducting state, SC1. **b**, Magnetoresistance of the $1T'$-$WS_2$ device when the magnetic field is oriented perpendicular to the *bc*-plane, plotted for different pressures. The superconducting transition shifts to lower fields upon increasing pressure, disappearing at $P = 1.15$ GPa and reappearing at 1.8 GPa, with a lower critical field when compared to SC1. **c**, Magnetoresistance of the $1T'$-$WS_2$ device when the magnetic field is oriented along the *c*-axis of the crystal, shown for different pressures. Akin to the data in panel **b**, the superconducting transition shifts to lower fields with increasing pressure in SC1 state. However, in contrast to the behavior observed for fields perpendicular to the *bc* plane, the superconducting transition in the SC2 state occurs at a similar temperature when compared to the SC1 state for fields oriented along the *bc*-plane. All data in panels **b** and **c** were collected at $T = 0.3$ K.

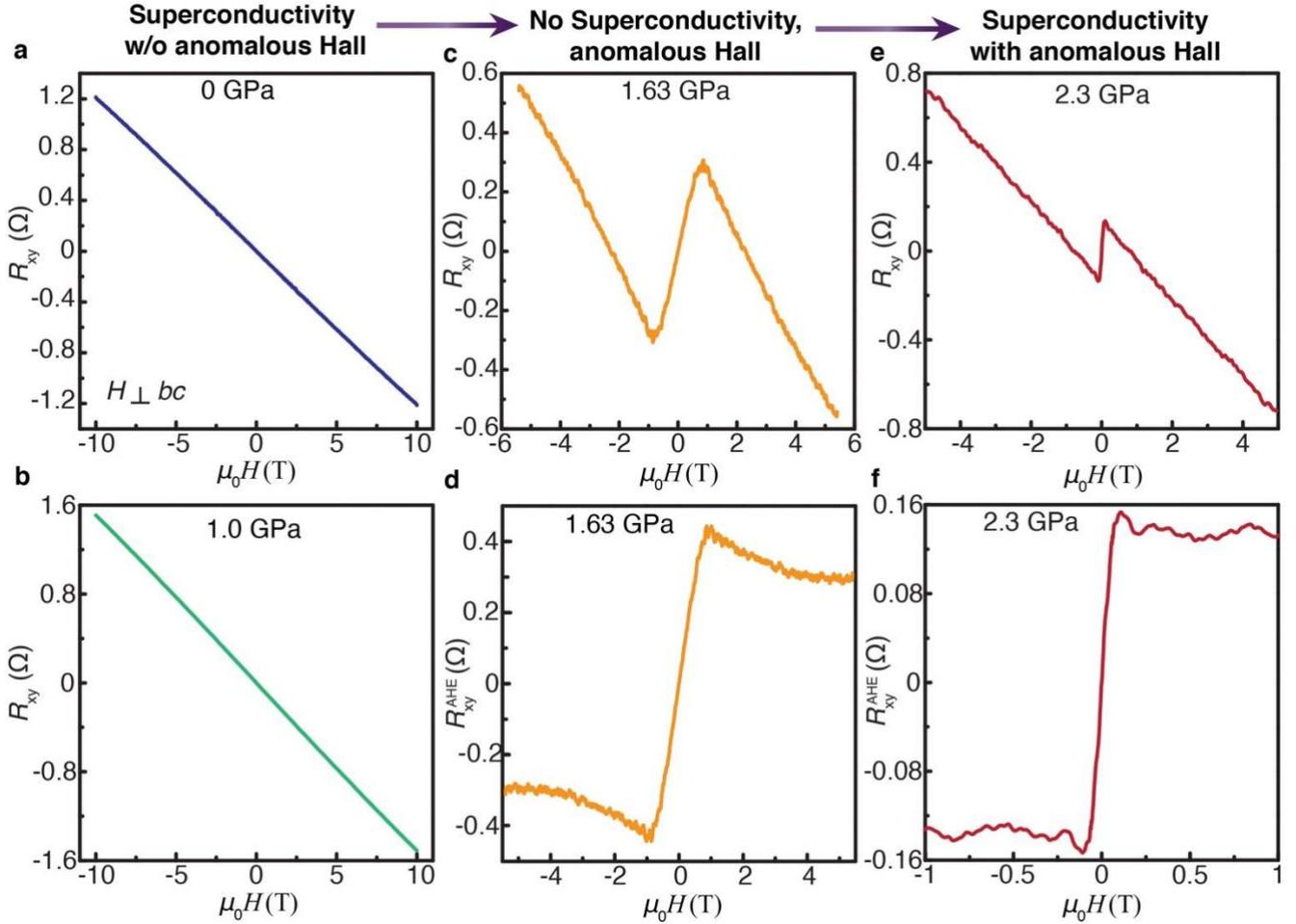

**Fig. 3: Emergence of the anomalous Hall effect upon the disappearance of the first superconducting state and its coexistence within the reentrant superconducting state.** Hall resistance ($R_{xy}$) at different pressures: $P =$ 0 GPa (panel **a**), 1.0 GPa (panel **b**), 1.63 GPa (panel **c**), and 2.3 GPa (panel **e**), plotted as a function of the magnetic field with fields oriented perpendicularly to the *bc*-plane. In panels **a** and **b**, $R_{xy}$ shows a linear response as a function of the field, displaying no anomalous Hall effect. However, at $P = 1.63$ GPa (panel **c**) and 2.3 GPa (panel **e**), along with a reasonably linear Hall effect, an anomalous component emerges at low fields. **d**, **f**, Anomalous Hall resistance, $R_{xy}^{AHE}$, obtained after subtracting the linear-in-field component of $R_{xy}$, plotted as a function of magnetic field for $P = 1.63$ GPa (panel **d**) and 2.3 GPa (panel **f**). Superconductivity is absent at $P = 1.63$ GPa where the anomalous Hall effect is more pronounced. Conversely, in the reentrant superconducting state at 2.3 GPa, superconductivity coexists with the relatively subdued anomalous Hall effect. All the data were acquired at $T = 10$ K.

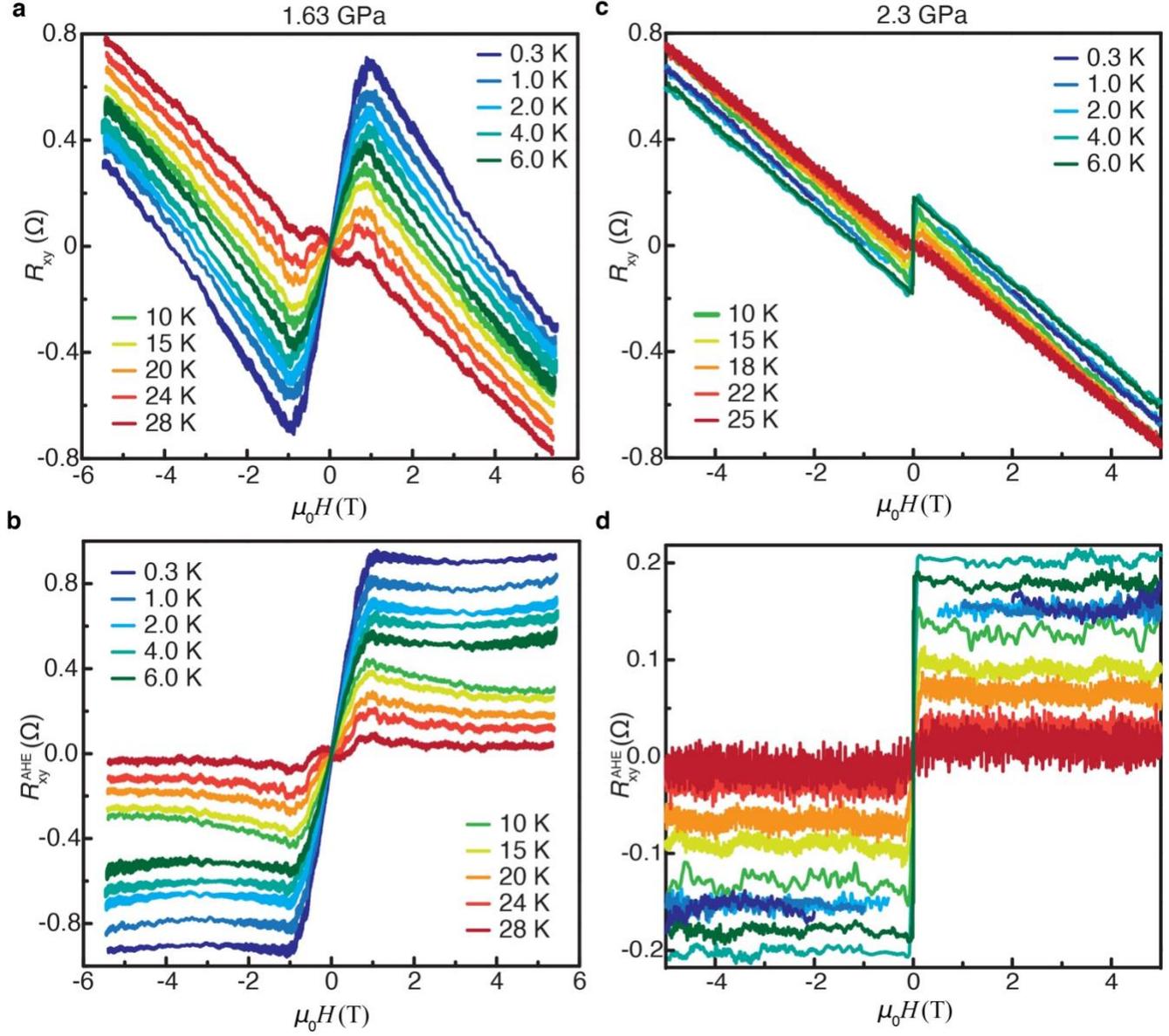

**Fig. 4: Temperature dependence of the anomalous Hall effect. a,** Hall resistance ($R_{xy}$) at $P = 1.63$ GPa, as a function of the magnetic field perpendicular to the *bc*-plane, for different temperatures. An anomalous Hall component is evident, and this component weakens as the temperature increases. **b,** Anomalous Hall resistance, obtained by subtracting the linear-in-field component of $R_{xy}$, as a function of the magnetic field at $P = 1.63$ GPa, and for different temperatures. The anomalous Hall effect diminishes with rising the temperature, and becomes almost entirely suppressed at $T = 28$ K. **c,** $R_{xy}$ at $P = 2.3$ GPa, where the reentrant superconductivity is present, as a function of the magnetic field perpendicular to the *bc*-plane, for different temperatures. The anomalous Hall component also weakens with increasing temperature. **d,** Anomalous Hall resistance, $R_{xy}^{AHE}$, obtained by subtracting the linear-in-field contribution from $R_{xy}$, as a function of magnetic field at $P = 2.3$ GPa, and for different temperatures. The anomalous Hall effect diminishes as the temperature rises and eventually becomes negligible at $T = 25$ K. Notably, the value of saturated $R_{xy}^{AHE}$ at $P = 2.3$ GPa in high magnetic fields (above $H_{c2}$) and below $T_c$ appears to slightly decrease when compared to its immediate higher-temperature value. This behavior contrasts with the one observed at $P = 1.63$ GPa, where the sample is no longer superconducting, and the saturated value of $R_{xy}^{AHE}$ continues to increase as the temperature is lowered. This observation suggests a possible competition between superconductivity and the anomalous Hall effect in the reentrant superconducting state at $P = 2.3$ GPa.

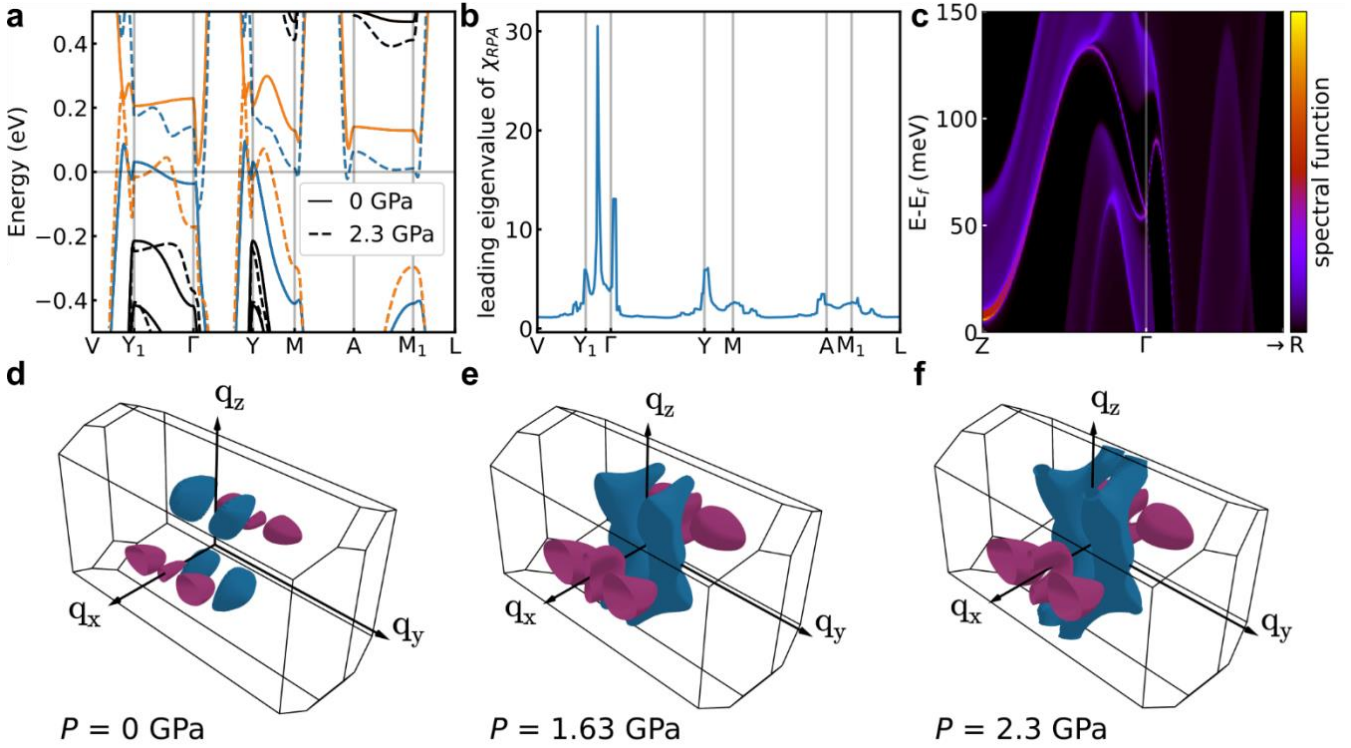

**Fig. 5. Pressure-tuned topological and Lifshitz transitions revealed from electronic band structure and susceptibility calculations. a**, Band structure as obtained from first principles calculation at two values of pressure $P = 0$ GPa, and 2.3 GPa, marked by solid and dashed lines, respectively. The band structure is shown along a path in the reciprocal space (see Extended data Table 2), and the color of the bands denotes the band inversion driving the change of topology (symmetry indicators $z_{2w1} = 1$, $z_{2w2} = 1$, $z_{2w3} = 0$, $z_4 = 0$ $P = 0$ GPa and $z_{2w1} = 0$, $z_{2w2} = 0$, $z_{2w3} = 0$, $z_4 = 3$ for $P = 2.3$ GPa). **b**, Leading eigenvalue of the random phase approximation (RPA) susceptibility evaluated along the same path in momentum space as in panel **a** for $P = 2.3$ GPa and a Hubbard interaction of $U = 6.5$ eV. The leading divergence discussed in the main text arises along the $Y_1$-$\Gamma$ line. **c**, Surface spectral function for a slab at $P = 2.3$ GPa, with a surface cone arising at $\underline{\Gamma}$ located at ~ 50 meV. The coordinates of $\underline{Z}$, $\underline{\Gamma}$ and $\underline{R}$ in terms of the basis vectors of the reciprocal cell in the semi-infinite slab geometry are (0,1/2), (0,0) and (1/2,1/2), respectively. **d**-**f**, Fermi surface as obtained from the first principles calculations at **d**, $P = 0$ GPa, **e**, $P = 1.63$ GPa, and **f**, $P = 2.3$ GPa in the first Brillouin zone. Blue (purple) Fermi surface sheets mark the electron-like (hole-like) pockets. In panels **e** and **f**, the Lifshitz transition fusing electron-like pockets and extending over Brillouin zone boundaries, respectively, is visible. Additionally, in panel **f**, a hole in the hole-like pocket along $q_x$ emerges.

## Methods:

### I. Synthesis of 1$T'$-WS$_2$ single crystals:

Initially, K$_2$WO$_4$ and S powders with a molar ratio of 1:4 were meticulously mixed and ground in a mortar with a total weight of 1.0 g. The resulting homogeneous mixture was then sealed in an evacuated quartz ampoule, which was then heated in a box furnace. The temperature was gradually elevated to 500 °C in approximately 6 hours, and then maintained for another 90 hours. The precursors were obtained after the furnace was turned off and naturally cooled down to room temperature. Next, 500 mg of the prepared precursor were placed into an alumina crucible within a quartz tube, which was cleaned and purged with H$_2$/Ar mixed gas (20% H$_2$/80% Ar) for 15 minutes 3 times. The crucible was then placed into the preheated furnace at 750 °C for further reaction. The reaction was maintained for 10 hours before rapidly cooling the quartz tube to room temperature. After that, the resulting product was thoroughly washed with Milli-Q water until the pH value of the suspension reached 7-8, and it was subsequently stored in Milli-Q water for 24 hours. To remove any remaining potassium residue, the crystals were transferred to an I$_2$ acetonitrile solution for another 24 hours. Finally, the 1$T'$-WS$_2$ crystals were obtained after twice washing with Milli-Q water followed by drying in a vacuum oven at room temperature.

We follow the nomenclature used in ref. [21] by classifying our crystals as 1$T'$-WS$_2$ (crystal structure is shown in Fig. 1**a**). The crystal structure file can be found in ref. [21]. We refer the reader to the supplementary Information for a detailed structural study of our 1$T'$-WS$_2$ crystals.

### II. Device fabrication and characterization:

We utilized a polydimethylsiloxane (PDMS) stamp-based mechanical exfoliation technique to fabricate atomically thin 1$T'$-WS$_2$ devices. The sample contacts on the silicon substrates were patterned with a 280 nm layer of thermal oxide using electron beam lithography, followed by chemical development and metal deposition (5 nm Cr/35 nm Au). The freshly exfoliated 1T′-WS$_2$ flakes were obtained from bulk single crystals and placed on PDMS stamps. To ensure uniformly thick samples with good geometry, we examined them under optical microscope before transferring them onto SiO$_2$/Si substrates with pre-patterned Cr/Au electrodes. Note that, we identified the thickness of the flakes by optical contrast, which is a commonly used method for the sensitive samples, see, e.g., ref. [36]. As widely known in the field of two-dimensional materials, samples with various thicknesses show different optical contrast[37]. Therefore, we characterized the thickness of the samples using atomic force microscopy and established the corresponding relationship between optical contrast and thickness beforehand. To preserve the intrinsic properties of the compound and minimize environmental effects, we encapsulated the samples with hexagonal boron nitride (h-BN) thin films, with thicknesses ranging from ~10 to ~30 nm. This encapsulation ensured that the samples on the devices were protected from direct exposure to air. All sample fabrication processes were performed in a glovebox with a gas purification system maintaining the environment at low levels of O$_2$ and H$_2$O (<1 ppm). For imaging, we used an Olympus BX 53M microscope to capture optical images, and AFM images were taken with a Bruker Dimension Icon3 in tapping mode.

### III. Pressure-dependent magnetotransport measurements:

Measurements at variable pressure were conducted using a standard oil-based piston pressure cell. The 1$T'$-WS$_2$ devices were securely mounted to the pressure cell using electrically insulating epoxy. To create hydrostatic pressure, the pressure cell was filled with Daphne 7575 oil, a suitable hydrostatic fluid. Once assembled, the pressure cell was placed in a hydraulic press, and pressure was applied by a piston fed through a hole in the threaded top screw of the cell. When the desired pressure was achieved, the top screw was clamped to lock in the pressure. For pressure measurements at low temperatures (around $T = 10$ K), a small ruby chip fixed to the tip of a fiber optic was utilized. The low-temperature and variable pressure magnetotransport measurements were carried out in a $^3$He variable temperature insert at a base temperature of 0.3 K, with magnetic fields up to 18 T. These measurements were conducted at the National High Magnetic Field Laboratory in Tallahassee, Florida, USA. As pressure can only be varied at room temperature, the samples underwent different thermal cycles during the measurements at different pressures. To ensure the reliability of the results, multiple devices were prepared and measured.

### IV. Extended Hall effect measurements:

This section covers the supplementary Hall effect data, commencing with an exploration of the magnetic field dependence of the Hall effect, conducted at $T = 10$ K across a range of pressures extending from $P \geq 1.15$ GPa. Extended Fig. 3 illustrates these findings, highlighting the emergence of the anomalous Hall effect within this pressure range. As depicted in Extended Fig. 3**a,b**, the anomalous Hall effect at $P = 1.15$ GPa (where no superconducting state is observed at low temperatures) exhibits greater prominence compared to the $P = 1.8$ GPa case (where superconductivity reemerges at low temperatures) demonstrated in Extended Fig. 3**c,d**. This observation implies that the presence of superconductivity weakens the anomalous Hall effect, suggesting a potential competition between superconductivity and a possible magnetic order.

### V. First-principles and RPA calculations

Full relativistic structure relaxations were performed via the Vienna *ab initio* simulation package[38] (VASP) for different values of hydrostatic pressure. Lattice parameters and atomic positions were optimized until the forces acting on ions were smaller than 0.01 eV/Å along all cartesian directions. The plane-wave basis cutoff was set to 550 eV, and partial occupations of states were determined based on the 1$^{st}$ order Methfessel-Paxton[39] method with a 0.2 eV smearing width. The evolution of lattice parameters with pressure is shown in Extended Fig. 5. The Full Potential-Local Orbital code[40] (FPLO) was employed to calculate the band structures of the relaxed crystal lattices, which turn out to be identical to the spectra computed with VASP. The Perdew-Burke-Ernzerhof[41] (PBE) exchange-correlation functional within the General Gradient Approximation (GGA) was employed in all VASP and FPLO calculations, and the Brillouin zone was sampled with a $12 \times 12 \times 12$ mesh in self-consistent calculations.

FPLO was employed to construct Wannier models for the structures at $P = 0$ GPa, 1.63 GPa, and 2.3 GPa. W atoms' *d*-orbitals and S atoms' *p*-orbitals served as localized basis for these calculations. Only hopping parameters to states within a radius of 35 Bohr units and with absolute value larger than 0.001 eV were preserved in the models. The FPLO parameters defining the energy windows adopted for each species of orbitals are written in Extended data Table 1. A comparison between *ab initio* bands and the spectrum of tight-binding models is shown in Extended Fig. 6.

The calculation of the surface spectral function in a semi slab geometry was performed via the interface pyfplo distributed with FPLO. The surface of the semi slab was considered perpendicular to the stacking direction. The energy range was sampled with 400 points, and every segment connecting two high-symmetry points in the BZ was divided into 500 points. The parameter controlling the number of layers was chosen as to achieve a penetration depth of four $WS_2$ monolayers.

The topology of the system was diagnosed within the framework of topological quantum chemistry[42]. According to this formalism, a material hosts a topological phase if the wave functions in its valence bands do not transform as a band representation under space-group symmetries, *i.e.,* if their transformation is not identical to that of Bloch waves constructed from maximally localized Wannier functions consistent with the space group. This criterion can be applied by checking if the irreducible representation of valence bands at maximal *k*-points of the BZ matches those of a band representation: if they do not match, the material is topological. In this work, the software IrRep[43] has been employed to calculate the irreducible representations of valence bands in the distinct phases observed in terms of pressure. In both phases, the irreducible representations of the lowest 36 bands — where 36 is the number of electrons in the primitive unit cell — do not match with those of any band representation of the space group, hence they are topological. Then, the software Check Topological Mat. from the Bilbao Crystallographic Server[44,45] has been applied to calculate the symmetry indicators of each phase, as well as to classify its topology.

The set of symmetry-based indicators[46] of topology in space group C2/m consists of three weak indices $z_{2wi}$ (with i = 1,2,3), each defined mod. 2, and a $z_4$ indicator defined mod. 4. This set of indicators can be represented as ($z_{2w1}$, $z_{2w2}$, $z_{2w3}$, $z_4$). Notably, an odd value for the $z_4$ index indicates a strong topological phase. In fact, the strong topological phase identified via DFT at high hydrostatic pressure values is, indeed, classified by the values ($z_{2w1} = 0$, $z_{2w2} = 0$, $z_{2w2} = 0$, $z_4 = 3$) of indicators.

In contrast, the topological crystalline phase observed at low hydrostatic pressure is classified by the values ($z_{2w1} = 1$, $z_{2w2} = 1$, $z_{2w2} = 0$, $z_4 = 0$). This corresponds to a weak topological phase[46] protected by the primitive translations **p₁** and **p₂**, which should exhibit an even number of Dirac cones on surfaces preserving these translation symmetries. Moreover, the weak phase is enriched by features characteristic of either mirror-Chern or hourglass topological insulators[47]. Refining the classification further requires a detailed analysis of the electronic structure on various terminations and finite geometries, which lies beyond the scope of this work.

In the mirror-Chern scenario, the mirror plane protects the additional topology, with the mirror-Chern number **C_m** = 2 serving as an additional invariant. Consequently, the crystal would show 2 (mod. 4) Dirac cones on a surface preserving the mirror and 2 (mod. 2) helical one-dimensional modes on a finite geometry preserving the two-fold rotation symmetry.

Alternatively, in the hourglass case, the additional topology is protected by the reflection with respect to the plane at **c₂**/4 with glide vector -**c₁**/2, and it is expected to feature hourglass fermions on surfaces preserving the glide, as well as one-dimensional helical modes in a cylinder geometry preserving the associated screw rotation.

From the tight-binding models, we calculate the interacting susceptibility in the random phase approximation (RPA) in the particle-hole channel[48], given by

$$\chi_{RPA}(q) = \frac{\chi^0(q,0)}{1-\chi^0(q,0)U(q)}.$$

Here $\chi_{RPA}(q)$, the bare susceptibility $\chi^0(q)$ and Hubbard interaction $U(q)$ are to be understood as matrices in both orbitals and spin. We employ a local intra-orbital Hubbard interaction of uniform strength $U$. The orbital resolved bare susceptibility in frequency and momentum space is given by

$$\chi^0_{o_1 o_2 o_3 o_4}(q,i\omega_n) = \frac{1}{N}\sum_{k,\mu,\nu} a_\mu^{o_4}(k) a_\mu^{o_2 *}(k) a_\nu^{o_1}(k+q) a_\mu^{o_3 *}(k+q) \frac{n_F(E_\mu(k)) - n_F(E_\nu(k+q))}{i\omega_n + E_\mu(k) - E_\nu(k+q)},$$

where $o_i$ are combined spin and orbital indices, $\mu/\nu$ denote the bands with energy eigenvalues $E_\mu(k)$ and $n_F(\epsilon)$ is the Fermi distribution function. The transformation between orbital and band space is given by the eigenvectors $a_\mu^{o_i}(k)$ encoding the weight of the orbital and spin configuration $o_i$ on band $\mu$. Due to spin-orbit coupling in the system, bands will in general not be spin polarized.

The divergence of the leading eigenvalue of $\chi_{RPA}$ at wave vector $q_{max}$ signals a fermi-surface instability towards either charge or spin order with the corresponding ordering vector. To distinguish these cases, we decompose the spin dependence of the interacting susceptibility in terms of Pauli matrices

$$\chi_{RPA}^{s_1 s_2 s_3 s_4}(q_{max}) = \chi_{RPA}^{\rho\sigma}(q_{max}) \sigma_{s_1 s_3}^{\rho} \sigma_{s_2 s_4}^{\sigma},$$

where $s_i$ is the spin part of the combined index $o_i$ and $\rho/\sigma = \{0, x, y, z\}$, where the zeroth Pauli matrix is defined to be a unit matrix, corresponding to the charge sector. We find that the dominant eigenvalues of $\chi_{RPA}^{\rho\sigma}(q_{max})$ are found for $\rho/\sigma = \{x, y, z\}$, rendering the instabilities magnetic in nature.

## VI. Electron-phonon coupling calculations

To investigate the electron-phonon coupling of $1T'$-WS$_2$ as a function of pressure, we conducted density functional theory calculations utilizing the QUANTUM ESPRESSO package[49]. The local density approximation (LDA)[50] with projected augmented-wave (PAW) method[51] in PSLIBRARY[52] and a plane waves cutoff of 60 Ry were used. A **k** mesh of $9 \times 9 \times 9$ points with a Methfessel-Paxton[37] smearing of 0.02 Ry was used. The structure was relaxed until the residual forces were less than 0.136 meV/Å for each pressure. The dynamical matrices and perturbation potentials are computed within density functional perturbation theory[53] on a $4 \times 4 \times 4$ **q** mesh in the absence of SOC. The EPW code[54,55] was employed to interpolate the electron-phonon matrix element onto $30 \times 30 \times 30$ **k** and **q** meshes to calculate the isotropic Eliashberg spectral function and electron-phonon coupling strength. Wannier functions[56] were constructed with W $d$ orbitals and S $p$ orbitals on a $8 \times 8 \times 8$ **k** mesh. The Dirac delta functions are replaced by Lorentzians of widths 25 meV and 0.05 meV for electrons and phonons, respectively.

Extended Fig. 7 depicts the phonon dispersions of $1T'$-WS$_2$ under pressures of 0 and 2 GPa. The calculations suggest that $1T'$-WS$_2$ gets stiffer under pressure. Consequently, there is no sign of phononic instability in $1T'$-WS$_2$ as a function of pressure. In Extended Fig. 8, the phonon density of states (DOS) $F(\omega)$ for $1T'$-WS$_2$ is depicted under pressures of 0, 1, and 2 GPa. The $F(\omega)$ profile notably shift towards higher phonon frequencies with increasing pressure, attributed to phonon hardening induced by a reduction in the lattice constant. The isotropic Eliashberg spectral functions $\alpha^2 F(\omega)$ and the cumulative electron-phonon coupling strength $\lambda(\omega)$ are presented in Extended Figs. 8**b** and 8**c**, respectively. Considering that $\lambda(\omega) = 2\int \alpha^2 F(\omega)\omega^{-1} d\omega$ is inversely proportional to $\omega$, the contribution of the low phonon frequency regime of $\alpha^2 F(\omega)$ to the total $\lambda$ is more substantial than that of the high-frequency regime. In Extended Fig. 8**b**, a discernible reduction in the intensity of $\alpha^2 F(\omega)$ is observed in the low-frequency range spanning from 0 to 15 meV as the pressure increases. Consequently, the intensity of $\lambda$ experiences a rapid decrease within this frequency range. Specifically, our calculations reveal that the total λ decreases from 0.528 to 0.446 as the pressure increases from 0 to 2 GPa.

## VII. Spin susceptibility calculations

To understand the effect of pressure on the suppression of superconductivity and enhancement of anomalous signatures in Hall response, we investigate the role of spin fluctuations. Specifically, we calculated the spin susceptibility by considering the effect of the external magnetic field on the $1T'$-WS$_2$ via Zeeman interaction. To facilitate this approach within the density functional theory calculation, we considered the 'fixed-spin-moment' approach as described in ref.[57]. Here, we simulate the effect of the external magnetic field by putting constraints on the magnetic moment of W atoms and obtaining the total energy as a function of magnetic moment $m$. The expansion of the DFT total energy will take the form of $E(m) = a0 + a1\,m^2 + a2\,m^4 + a3\,m^6 + a4\,m^8 + a5\,m^{10} + \cdots$. The spin susceptibility, denoted as $\chi$, can be estimated from $\chi = \left(\frac{\partial^2 E}{\partial m^2}\right)^{-1}$. For small values of $m$, this can be approximated as $\chi = \frac{1}{a1}$, where $a1$ can be determined by fitting the density functional theory total energy to $E(m)$ as depicted in Extended Fig. 9 for three different pressures. We found the spin susceptibility to be 0.8110 eV/$\mu_B^2$, 1.0900 eV/$\mu_B^2$ and 1.0900 eV/$\mu_B^2$ for 0 GPa, 1 GPa, and 2 GPa, respectively. The observed increase in spin susceptibility indicates stronger spin fluctuations in this system with increasing pressure, yet no ferromagnetic instability by spin ordering.

**Methods Only References:**

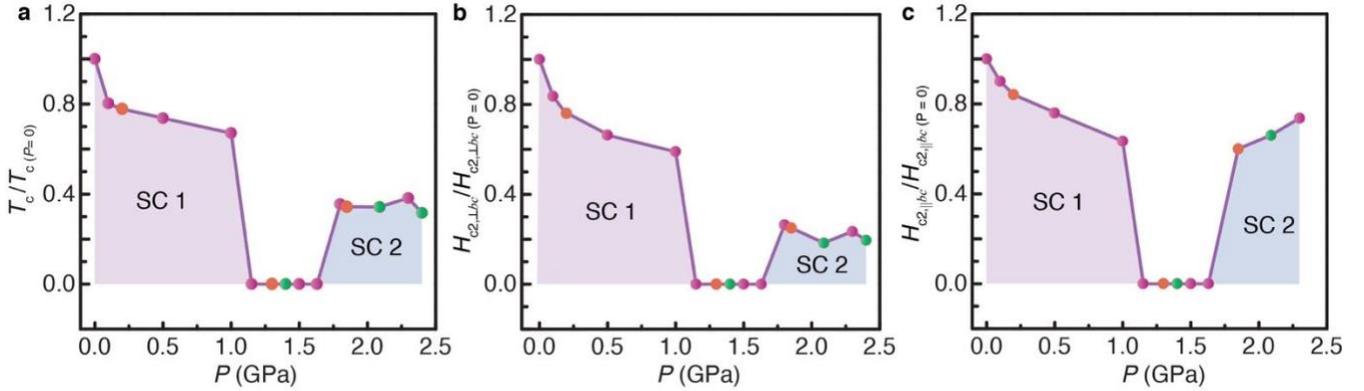

**Extended Fig. 1: Tunable superconducting states in few layered 1$T'$-WS$_2$.** Pressure dependence of $T_c$ (panel **a**), $\mu_0 H_{c2\perp bc}$ (panel **b**), $\mu_0 H_{c2||bc}$ (panel **c**), respectively, normalized by the corresponding values collected under vacuum.

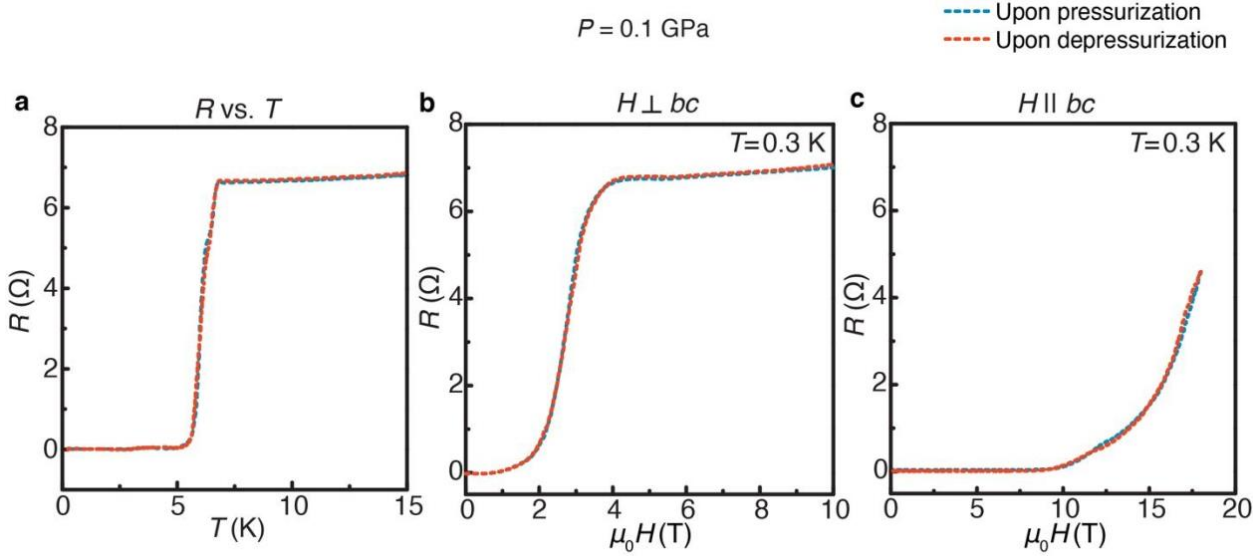

**Extended Fig. 2: Transport measurements on 1$T'$-WS$_2$ at a pressure of 0.1 GPa, conducted during both pressurization and depressurization cycles (starting from 2.3 GPa), demonstrating reversibility in the transport. a**, Temperature ($T$) dependence of the four-probe resistance ($R$) for the 1$T'$-WS$_2$ device **b**, Magnetoresistance of 1$T'$-WS$_2$ when the magnetic field is oriented perpendicularly to the $bc$-plane. **c**, Magnetoresistance of the 1$T'$-WS$_2$ when the magnetic field is oriented along the $c$-axis of the crystal. In all panels, measurements collected upon pressurization and depressurization show consistent, overlapping resistivity, indicating reversibility. Data in panels **b** and **c** were recorded at $T = 0.3$ K.

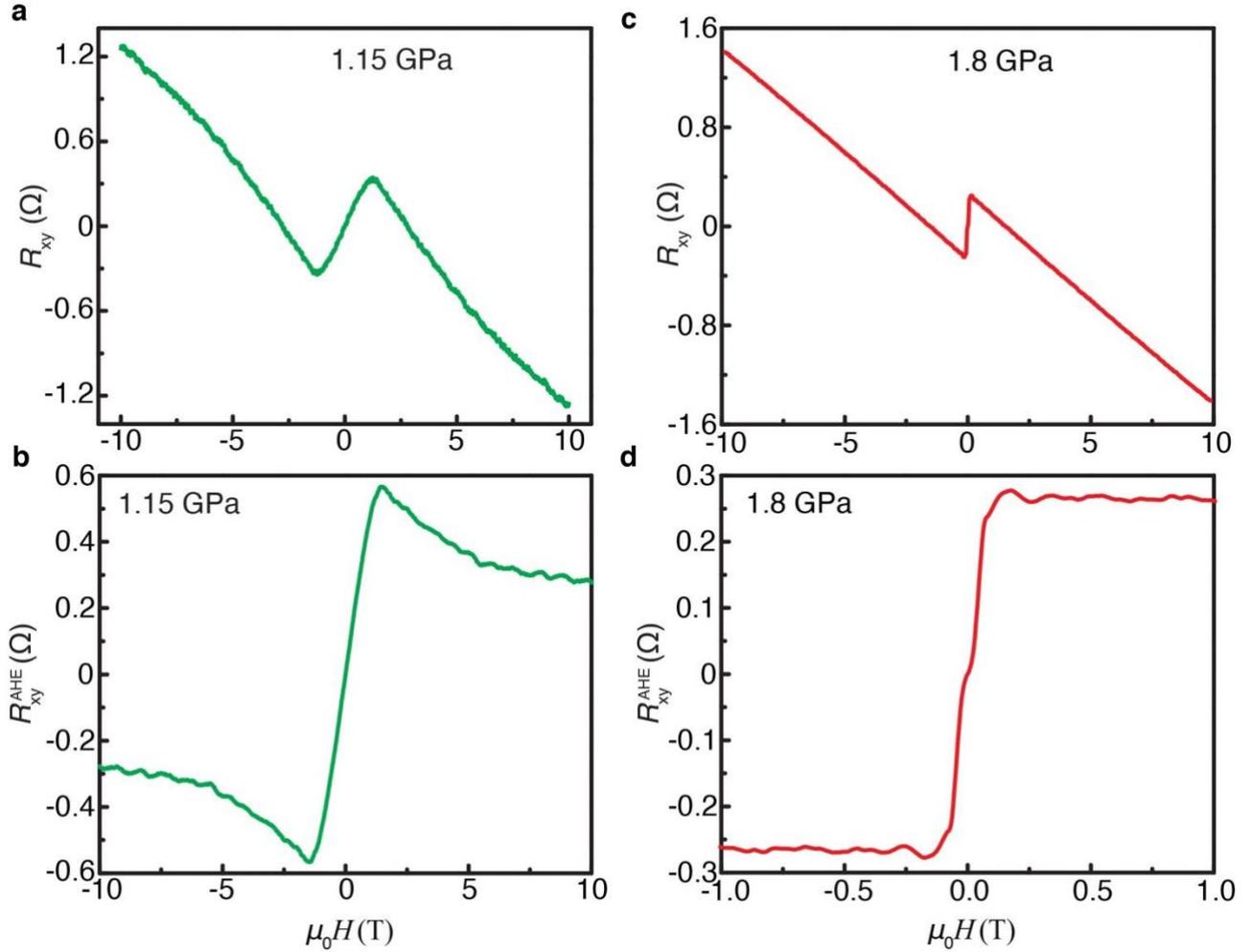

**Extended Fig. 3: Observation of the anomalous Hall effect at pressures $P \geq 1.15$ GPa**. **a**, Hall resistance ($R_{xy}$) at $P = 1.15$ GPa, precisely at the point of superconductivity disappearance, plotted against the perpendicular-to-$bc$-plane magnetic field. The presence of a distinct anomalous Hall component is evident. **b**, Anomalous Hall resistance, acquired by subtracting the linear-in-field portion of $R_{xy}$, depicted as a function of the magnetic field at $P = 1.15$ GPa. **c**, $R_{xy}$ at $P = 1.8$ GPa, where the reentrant superconductivity emerges, presented as a function of the magnetic field perpendicular to the $bc$-plane. The anomalous Hall component persists at 1.8 GPa. **d**, Anomalous Hall resistance, $R_{xy}^{AHE}$, obtained after subtracting from $R_{xy}$ the linear-in-field contribution, shown as a function of the magnetic field at $P = 1.8$ GPa. The anomalous Hall effect is more pronounced at $P = 1.15$ GPa where superconductivity is absent and relatively subdued at 1.8 GPa where the reentrant superconducting state coexists.

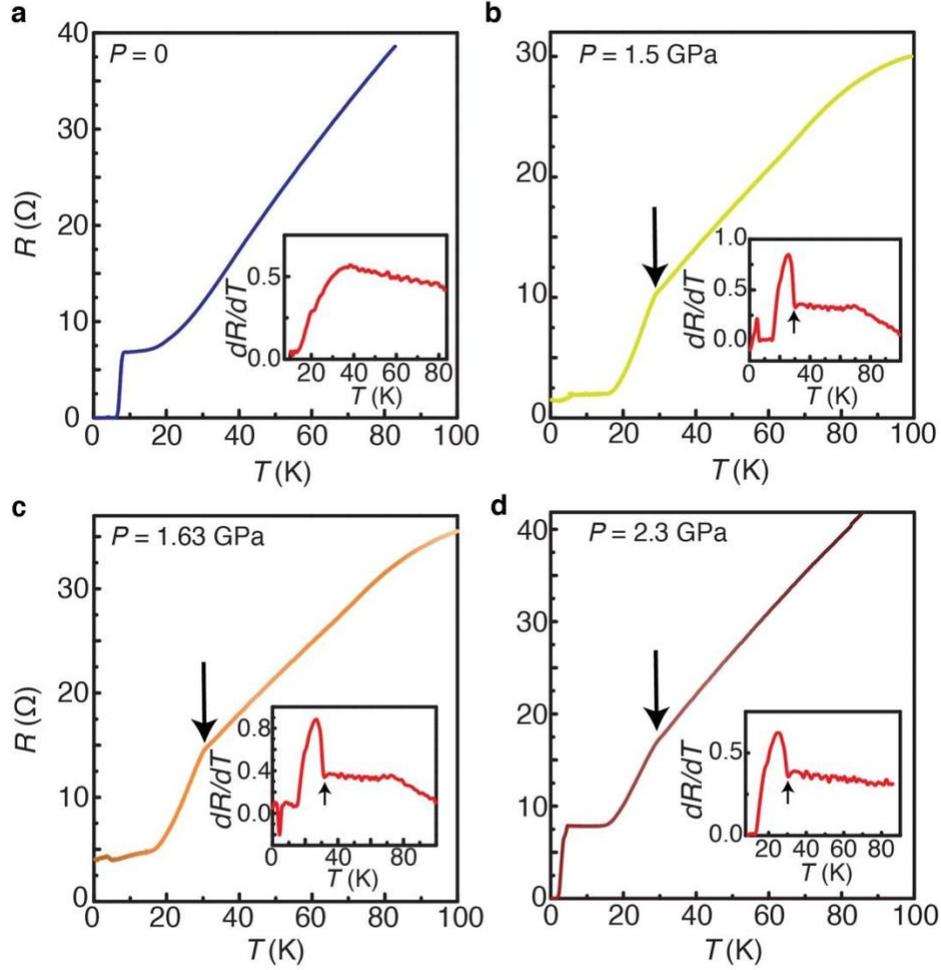

**Extended Fig. 4: Transport measurements under several pressures revealing a weak anomaly around $T = 30$ K that correlates with the disappearance of the first superconducting state. a-d**, Temperature ($T$) dependence of the four-probe resistance ($R$) for a $1T'$-WS$_2$ sample measured at different hydrostatic pressures ($P$). Insets: corresponding d$R$/d$T$ as a function of $T$. $1T'$-WS$_2$ exhibits metallic behavior at all pressures. At $P = 0$, $R$ as a function of $T$ shows no resistance anomalies (panel a). By contrast, at $P = 1.5$ GPa (panel **b**), 1.63 GPa (panel **c**), and 2.3 GPa (panel **d**), an anomaly near $T = 30$ K is present in the resistance. These anomalies become evident in the d$R$/d$T$ plots.

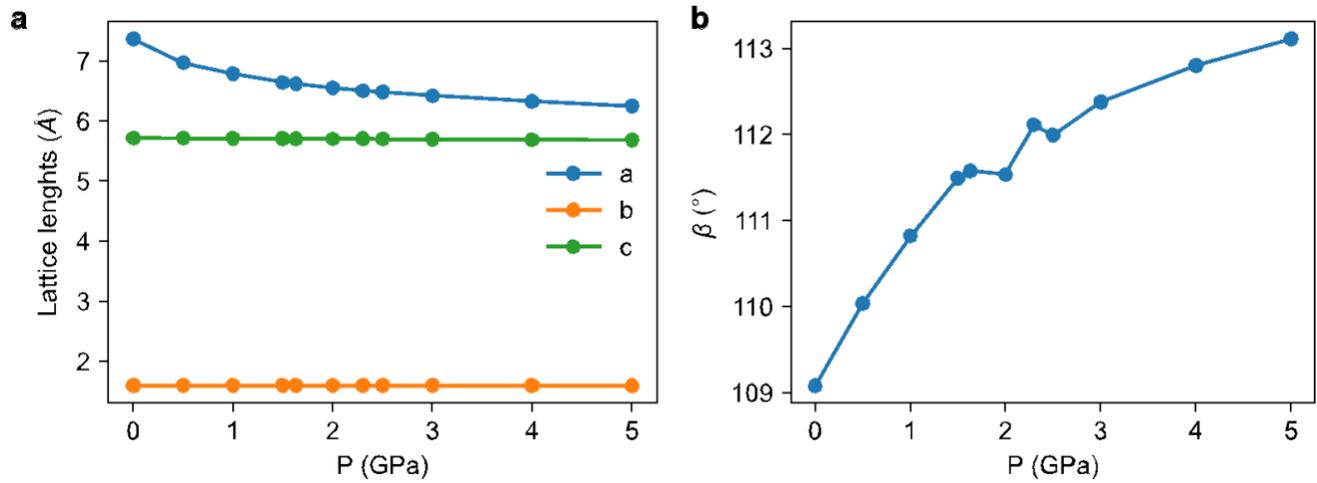

**Extended Fig. 5: Evolution of lattice parameters after the structural optimization via VASP. a,** Evolution of the lengths of the vectors defining the centered unit cell. **b,** Evolution of the monoclinic angle.

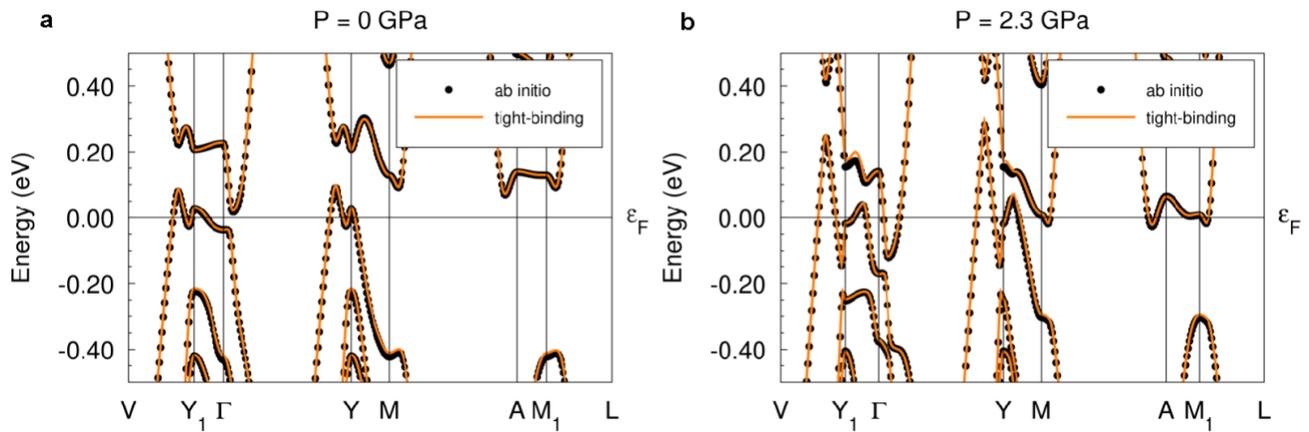

**Extended Fig. 6: Comparison between *ab initio* electronic band structures obtained with FPLO and the band structures from Wannierized tight-binding models. a,** $P = 0$ GPa. **b,** $P = 2.3$ GPa.

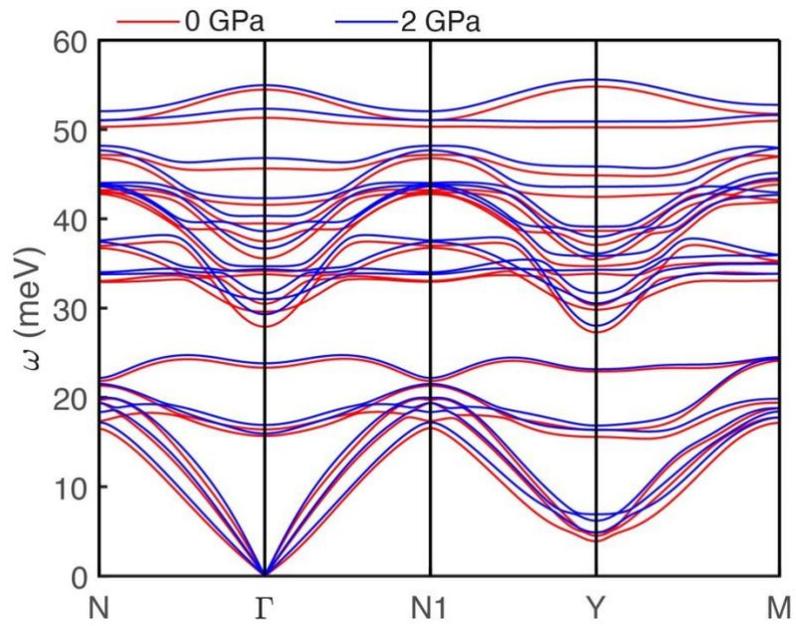

**Extended Fig. 7: Calculated phonon dispersions for 1$T'$-WS$_2$ at 0 GPa (red) and 2 GPa (blue).** The absence of imaginary frequencies confirms the dynamical stability of 1$T'$-WS$_2$ under 2 GPa.

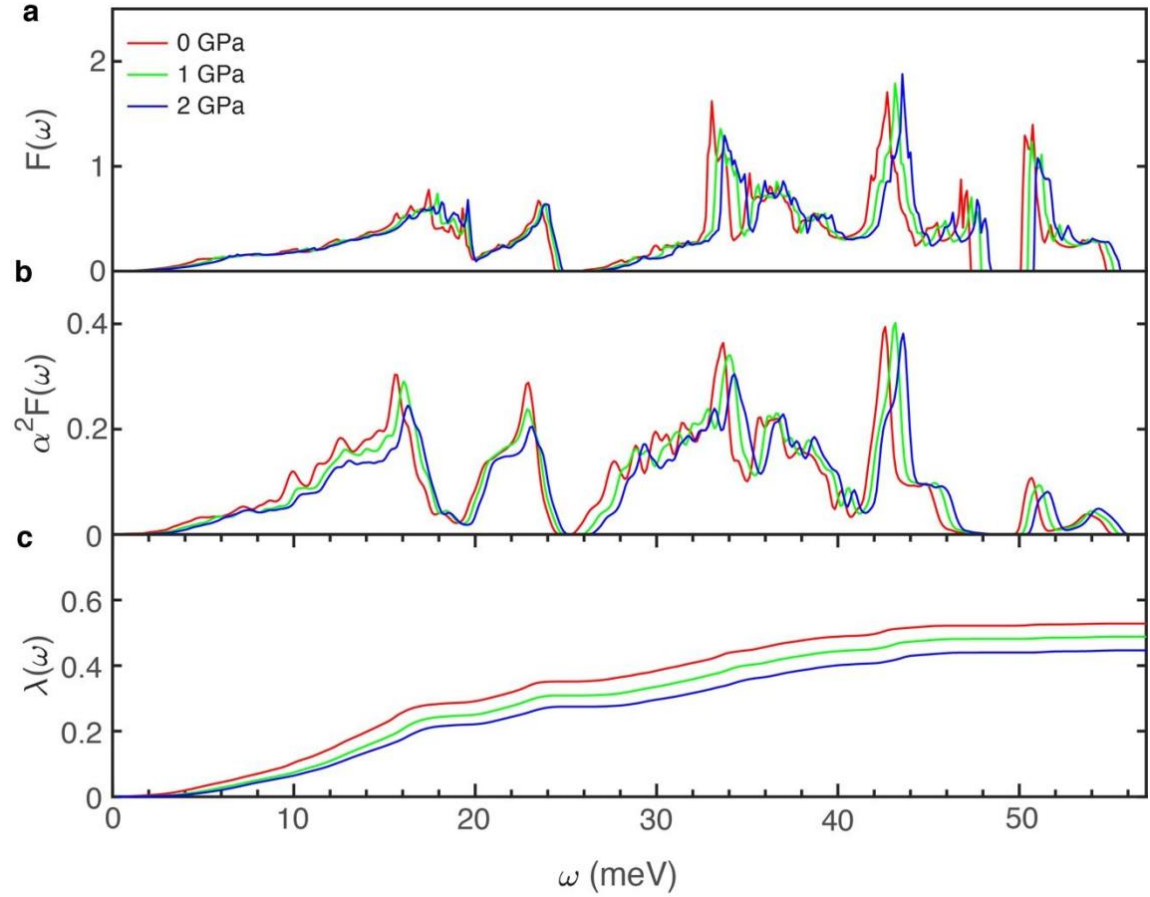

**Extended Fig. 8: Pressure dependence of the electron-phonon coupling. a,** Phonon density of states $F(\omega)$. **b,** Isotropic Eliashberg spectral function $\alpha^2 F(\omega)$. **c,** Cumulative electron-phonon coupling strength $\lambda(\omega)$. Pressure-dependent profiles of the electron-phonon coupling in $1T'$-WS$_2$ for 0 GPa (red), 1 GPa (green), and 2 GPa (blue) are represented by solid lines.

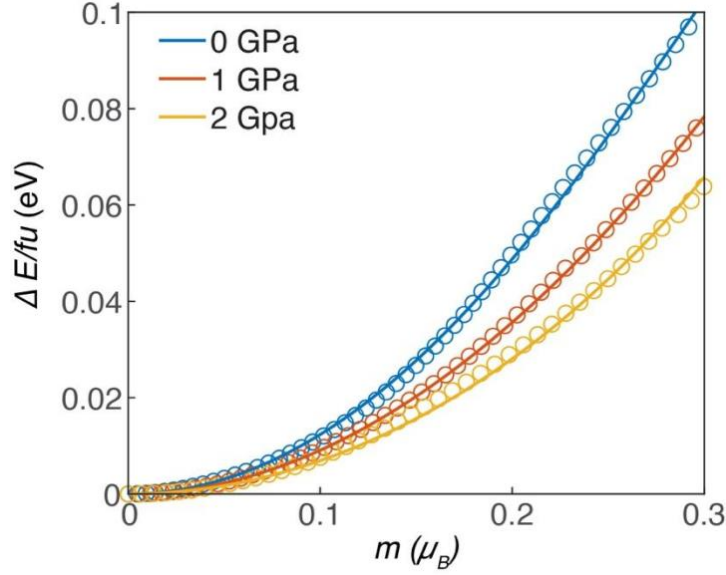

**Extended Fig. 9: Determining the pressure-dependent spin susceptibility.** Fitting of the density functional theory total energy to $E(m)$ under three distinct pressures (0 GPa, 1 GPa, and 2 GPa). The spin susceptibility values obtained are 0.8110 eV/$\mu_B^2$, 1.0900 eV/$\mu_B^2$ and 1.0900 eV/$\mu_B^2$ for 0 GPa, 1 GPa, and 2 GPa, respectively. The escalating spin susceptibility suggests intensified spin fluctuations in this system with increasing pressure.

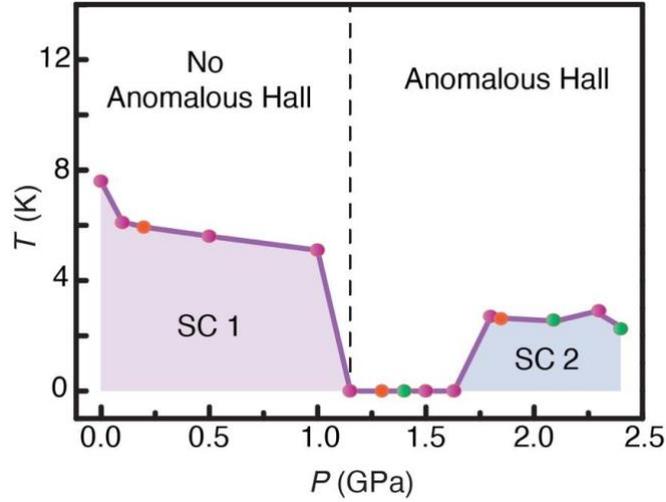

**Extended Fig. 10. Pressure-tunable electronic and superconducting states in atomically thin 1$T'$-WS$_2$.** Under hydrostatic pressure, our atomically thin 1$T'$-WS$_2$ sample undergoes a cascade of phase transitions. First, the superconducting state weakens progressively with increasing pressure, as is evident from the decrease of the $T_c$ and $H_{c2}$ values (as illustrated in Fig. 1), until it eventually disappears at 1.15 GPa. Below this critical pressure, the Hall effect maintains linearity as a function of the magnetic field, and no anomalous Hall effect is observed. However, at 1.15 GPa, right where superconductivity vanishes, an anomalous Hall effect emerges, suggesting a possible magnetic state. At $P = 1.8$ GPa, a 'Lazarus effect' occurs, with the re-emergence of superconductivity, albeit with significantly different properties. Intriguingly, the anomalous Hall effect persists within this reentrant superconducting state, pointing to the coexistence between possible magnetism and superconductivity or to an unconventional superconducting state.

**Extended data Table 1: FPLO parameters for the energy windows of orbitals in the tight-binding models.**

| Orbital type | emin (eV) | emax (eV) | delower (eV) | deupper (eV) |
|---|---|---|---|---|
| *d*-orbitals of W | -1.0 | -1.0 | 6.2 | 4.1 |
| *s*-orbitals of S | -1.0 | -1.0 | 6.2 | 8 |

**Extended data Table 2: Coordinates of high-symmetry *k*-points in the primitive setting of the unit cell.**

| V | $Y_1$ | Γ | Y | M | A | $M_1$ | L |
|---|---|---|---|---|---|---|---|
| (1/2, 0, 0) | (1/2, -1/2, 0) | (0, 0, 0) | (1/2, 1/2, 0) | (1/2, 1/2, 1/2) | (0, 0, 1/2) | (1/2, -1/2, 1/2) | (1/2, 0, 1/2) |

**Acknowledgments**

M.Z.H. group acknowledges primary support from the US Department of Energy, Office of Science, National Quantum Information Science Research Centers, Quantum Science Center (at ORNL) and Princeton University; STM Instrumentation support from the Gordon and Betty Moore Foundation (GBMF9461) and the theory work; and support from the US DOE under the Basic Energy Sciences programme (grant number DOE/BES DE-FG-02-05ER46200) for the theory and sample characterization work including ARPES. The work of M.I. was funded by the European Union NextGenerationEU/PRTR-C17.I1, as well as by the IKUR Strategy under the collaboration agreement between Ikerbasque Foundation and DIPC on behalf of the Department of Education of the Basque Government. M.I. thanks support to the Spanish Ministerio de Ciencia e Innovacion (grant PID2022-142008NB-I00). T.M. acknowledges funding by the Deutsche Forschungsgemeinschaft (DFG, German Research Foundation) through the Würzburg-Dresden Cluster of Excellence on Complexity and Topology in Quantum Matter – ct.qmat Project-ID 390858490- EXC 2147. L.B. is supported by DOE-BES through award DE-SC0002613. The National High Magnetic Field Laboratory (NHMFL) acknowledges support from the US-NSF Cooperative agreement Grant number DMR-DMR-2128556 and the state of Florida. We thank T. Murphy, G. Jones, L. Jiao, and R. Nowell at NHMFL for technical support. T.N. acknowledges supports from the European Union's Horizon 2020 research and innovation programme (ERC-StG-Neupert-757867-PARATOP). H.Z. thanks the support from ITC via the Hong Kong Branch of National Precious Metals Material Engineering Research Center (NPMM), the Research Grants Council of Hong Kong (AoE/P-701/20), the Start-Up Grant (Project No. 9380100) and grant (Project No. 1886921) from the City University of Hong Kong, and the Science Technology and Innovation Committee of Shenzhen Municipality (grant no. JCYJ20200109143412311). K.W. and T.T. acknowledge support from the JSPS KAKENHI (Grant Numbers 20H00354 and 23H02052) and World Premier International Research Center Initiative (WPI), MEXT, Japan. Z.L. acknowledges the start-up funding support (Project ID: 1-BE7U) from The Hong Kong Polytechnic University.